\documentclass[]{spie}  
\usepackage[]{graphicx,amssymb,amsmath}

\title{Mode-locked oscillators in the positive and negative dispersion regimes: scenarios of destabilization}

\author{Vladimir L. Kalashnikov and Evgeni Sorokin
\skiplinehalf Institut f\"{u}r Photonik, TU Wien, Gusshausstr.
27/387, A-1040 Vienna, Austria}

\authorinfo{Further author information: (Send correspondence to V.L.Kalashnikov) \\V.L.Kalashnikov:
E-mail: kalashnikov@tuwien.ac.at, Telephone: +43 1 588 01 387 43}

\begin{document}
  \maketitle

\begin{abstract}
We analyze the influence of spectrally modulated dispersion and loss
on the stability of mode-locked oscillators. In the negative
dispersion regime, a soliton oscillator can be stabilized in a close
proximity to zero-dispersion wavelength, when spectral modulation of
dispersion and loss are strong and weak, respectively. If the
dispersion is close to zero but positive, we observe \emph{chaotic}
mode-locking or a stable coexistence of the pulse with the CW
signal. The results are confirmed by experiments with a Cr:YAG
oscillator.
\end{abstract}



\section{INTRODUCTION}
\label{intro}  

Oscillators providing stable sub-100 fs pulses in the near-infrared
region around 1.5~$\mu$m are of interest for a number of
applications including infrared continuum generation
\cite{kalash1,kalash2} and high-sensitivity gas spectroscopy
\cite{mandon}. To date, the typical realization of such sources is
based on a femtosecond Er:fiber oscillator with an external
pulse amplification. A promising alternative to such
combination is a solid-state Cr$^{4+}$:YAG mode-locked oscillator
\cite{naumov,leburn}. Such an oscillator allows a direct diode
pumping and possesses the gain band providing the few-optical cycle
pulses.

However, attempts to increase the pulse energy in a Cr:YAG
oscillator is limited by its relatively small gain coefficient.
Because of the low gain, the oscillator has to operate with low
output couplers and, thereby, the intra-resonator pulse energy has
to be high. As a result, the instabilities appear
\cite{naumov,kalash3}. To suppress the instabilities in the negative
dispersion regime (NDR) a fair amount of the group-delay-dispersion
(GDD) is required. The resulting pulse is a relatively long soliton
with reduced peak power. Such a pulse is nearly transform-limited
and is not compressible. A remedy is to use the positive dispersion
regime (PDR), when the pulse is stabilized due to substantial
stretching (up to few picoseconds) caused by a large chirp
\cite{apol}. Such a pulse is dispersion-compressible down to few
tens of femtoseconds.

For both NDR and PDR, the oscillator will eventually become unstable at high power.
The main scenarios of the pulse destabilization have
been identified with the multipulsing in the NDR \cite{kalash3} and
the CW-amplification in the PDR \cite{kalash4}. It has been found,
that the higher-order dispersions (i.e., the frequency dependent
GDD) and losses significantly modify the stability conditions
\cite{kalash5,kalash6}. Hence, the study of the stability conditions
affected by both linear and nonlinear processes inherent in a
mode-locked oscillator remains an important task.

Here, we present a study of the destabilization mechanisms of a
Cr:YAG mode-locked oscillator, operating in both NDR and PDR. We put a
special emphasis on the influence of the spectral dependence of
GDD and losses on the oscillator stability.

\section{DESTABILIZATION OF A MODE-LOCKED OSCILLATOR IN THE PDR}\label{s1}

\begin{figure}
   \begin{center}
   \vspace{-5 mm}
   \includegraphics[height=8cm]{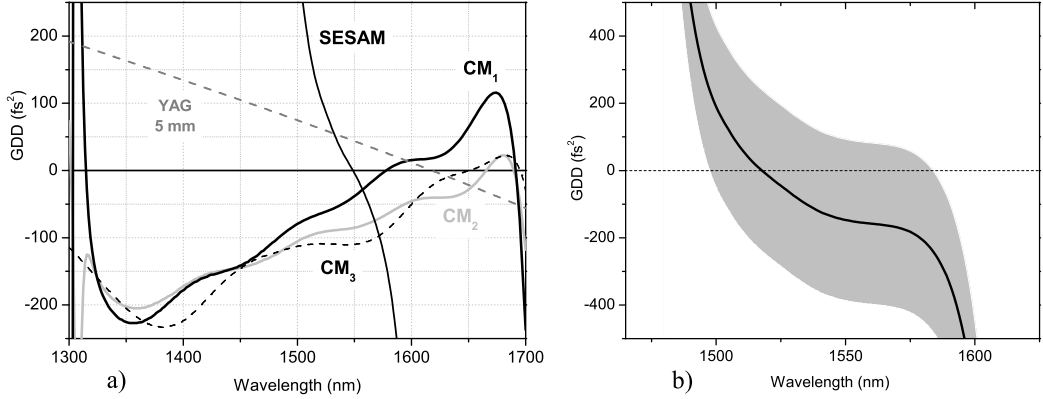}
   \vspace{-1 cm}
   \end{center}
   \caption[fig1]
   { \label{fig1}
a) GDD of three sets of chirped mirrors CM (as designed), the YAG
crystal, and the SESAM.  b) The net dispersion of the resonator of
Cr$^{4+}$:YAG oscillator. Black line: as designed, grey area -
uncertainty region due to the chirped mirrors.}
   \end{figure}

The Cr:YAG oscillator has been built on the basis of the scheme
published in Refs. \cite{kalash2,setup}. The mode-locking and the
dispersion control were provided by SESAM and chirped-mirrors (CMs),
respectively. The GDD of intra-resonator elements as well as the
net-GDD are shown in Fig. \ref{fig1}. As a result of the GDD
variation of the 51-layer CMs and the uncertainty of the SESAM
dispersion, the real net-GDD has some uncertainty, too (gray region
in Fig. \ref{fig1}, \emph{a}).

\begin{figure}
   \begin{center}
   \vspace{-5 mm}
   \includegraphics[height=8cm]{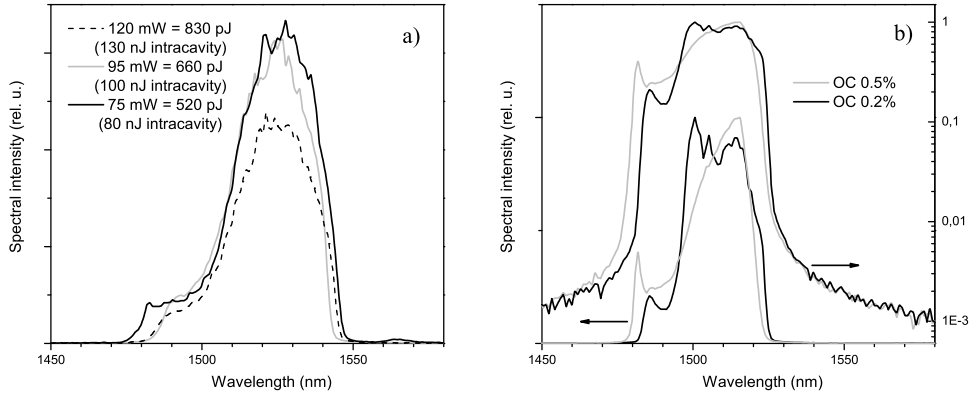}
   \vspace{-1 cm}
   \end{center}
   \caption[fig2]
   { \label{fig2}
a) Spectra of the Cr:YAG oscillator operating in the PDR at
different values of intracavity pulse energy. b) Spectra of the
Cr:YAG oscillator with different output couplers.}
   \end{figure}

Selection of the different CM combinations allows over- and under-compensation of the
dispersion. Selecting a 2CM$_1$+2CM$_1$ allows stabilizing the oscillator at the 144.5
MHz pulse repetition rate and 150 mW average output power. The
corresponding spectra shown in Fig. \ref{fig2} have truncated
profiles, that is typical for an oscillator operating in the PDR
\cite{kalash4}.

To study the stability limits of PDR, the numerical simulations
based on the nonlinear cubic-quintic complex Ginzburg-Landau model
\cite{akhmed} have been realized. The evolution of the slowly
varying field envelope $A(z,t)$ can be described in the following
way:

\begin{eqnarray} \label{eq1}
  \frac{{\partial A}}{{\partial z}} & = & \sigma \left( {\omega _0 } \right)A + \alpha \left(
{\omega _0 } \right)\tau ^2 \frac{{\partial ^2 A}} {{\partial t^2 }}
- \rho \left( {\omega  - \omega _0' } \right)A +  i\beta \left(
{\omega  - \omega _0 } \right)A + \left\{
   \left[ \kappa  - i\gamma \right]P - \kappa \varsigma P^2  \right\}A. \hfill
\end{eqnarray}

\noindent Here $z$ is the propagation distance normalized to the
cavity length $L_{cav}$ (i.e., the cavity round-trip number), $t$ is
the local time. The reference time frame moves with the pulse
group-velocity defined at the reference frequency $\omega_0$
corresponding to the gain maximum at $\lambda_0\approx$1.5 $\mu$m.
The term $\alpha(\omega_0) \tau^2 {{\partial ^2 A} \mathord{\left/
 {\vphantom {{\partial ^2 A} {\partial t^2 }}} \right.
 \kern-\nulldelimiterspace} {\partial t^2 }}
$ describes the action of the gain spectral profile in parabolic
approximation. Parameter $\alpha(\omega_0)=$0.028 is the saturated
gain coefficient at $\omega_0$, and it is close to the net-loss
value at this frequency. Parameter $\tau=$9.5 fs is the inverse gain
bandwidth. The term $\beta(\omega - \omega_0) A$ describes the
net-GDD action in the Fourier domain. The term $\rho(\omega -
\omega_0') A$ describes the action of the net loss spectral profile
in the Fourier domain. Frequency $\omega_0'$ corresponds to the
transmission minimum of the output coupler at
$\lambda_0'\approx$1.53 $\mu$m. Parameter $\gamma=$1.8 MW$^{-1}$
describes the self-phase modulation inside the active medium,
$\kappa=$0.05$\gamma$ is the self-amplitude modulation parameter,
$\zeta=$0.6$\gamma$ is the parameter defining saturation of the
self-amplitude modulation with power \cite{kalash4}.

Parameter $\sigma(\omega_0)$ is the difference between the saturated
gain $\alpha (\omega_0)$ and the net loss at the reference frequency
$\omega_0$. It was assumed, that this parameter depends on the full
pulse energy:$ \sigma \left( E \right) \approx \left.
{\frac{{d\sigma }}{{dE}}} \right|_{E = E^* } (E-E^*) = \delta\left(
{E/E^* - 1} \right)$, where $E^*$ corresponds to the full energy
stored inside an oscillator in the CW regime \cite{kalash4}.
Parameter $\delta \equiv \left. {\frac{{d\sigma }}{{dE}}} \right|_{E
= E^* } E^*$ equals to -0.03.

\begin{figure}
   \begin{center}
   \vspace{-5 mm}
   \includegraphics[height=8cm]{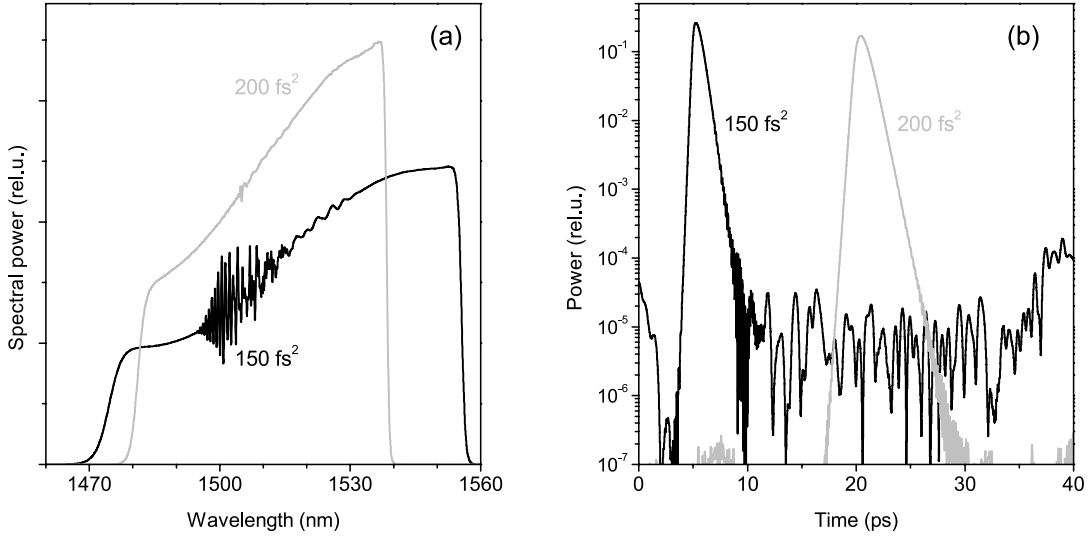}
   \vspace{-5 mm}
   \end{center}
   \caption[fig3]
   { \label{fig3}
(a) - Simulated spectra, (b) - intensity profiles in the positive dispersion regime. GDD corresponds to $\beta_2=$ 200
fs$^2$ (gray
  curves) and 150 fs$^2$ (black curves).}
   \end{figure}

It was found, that the GDD decrease in the PDR results in the
CW-amplification (see the black curve in Fig. \ref{fig3}). The
CW-amplification appears in the vicinity of the spectral net-loss
minimum, where the saturated net-gain becomes positive. The latter
occurs because gain saturation decreases with GDD approaching to
zero. Simultaneously, the spectrum gets broader, which
enhances the spectral losses and, thereby reduces the pulse energy.

\begin{figure}
   \begin{center}
      \vspace{-5 mm}
   \includegraphics[height=8cm]{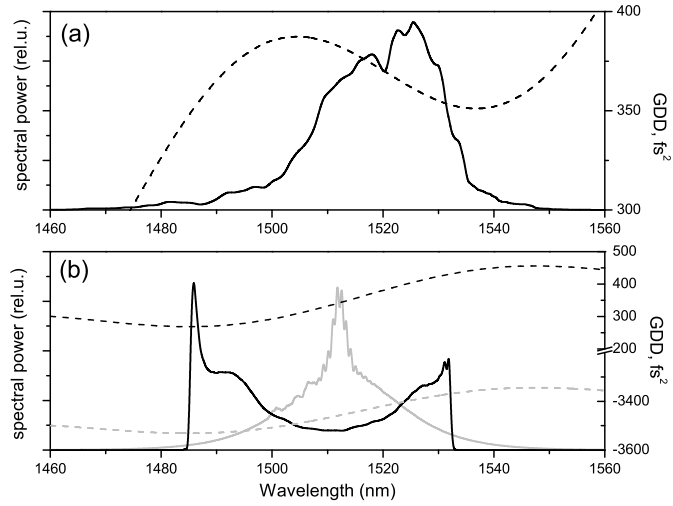}
      \vspace{-5 mm}
   \end{center}
   \caption[fig4]
   { \label{fig4}
Simulated spectra (solid curves), with different net-GDD (dashed
curves).
  $\lambda_0=$1.54
  $\mu$m, other parameters correspond to Fig. \ref{fig3}, the case without spectrally-dependent losses.}
   \end{figure}

When the GDD is spectrally dependent (i.e., there are the
higher-order dispersions), the scenario of destabilization with an
approaching of GDD to zero changes (Fig. \ref{fig4}, \emph{a}). In
this case, the chaotic oscillations of the peak power appear
(\emph{chaotic mode-locking} \cite{kalash5}). The spectrum edges
in such a regime become smoothed and the spectrum shifts to the
local GDD-minimum. It was found \cite{kalash5}, that the PDR
stability is improved in the vicinity of the local
minimum of the GDD (i.e., when the fourth-order dispersion is
positive). Fig. \ref{fig4}, \emph{b} shows the spectrum (gray curve)
corresponding to the lower on GDD stability border of PDR. The
spectrum is asymmetrical, ``M-shape'', and the spectrum maximum is
situated in the vicinity of the GDD local minimum.

\section{DESTABILIZATION OF A MODE-LOCKED OSCILLATOR IN THE NDR}\label{s2}

\begin{figure}
   \begin{center}
      \vspace{-5 mm}
   \includegraphics[height=6cm]{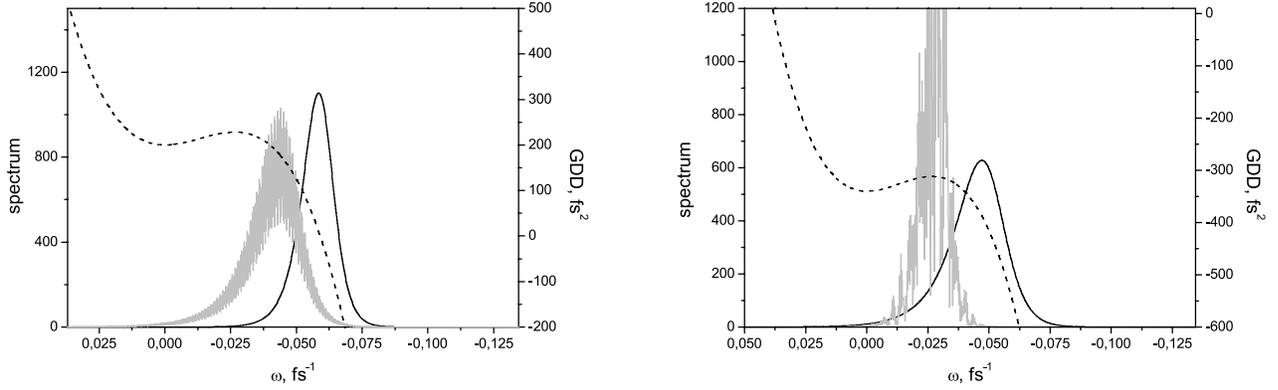}
   \vspace{-1 cm}
   \end{center}
   \caption[fig5]
   { \label{fig5}
GDD spectral profiles (dashed) and corresponding simulated
  pulse spectra without (black) and with (gray) spectral dependence of the losses. The losses are assumed to have a local minimum at 1.53 $\mu$m about -0.025 fs$^{-1}$ }
   \end{figure}

In the NDR and in the absence of higher-order dispersions, the pulse
can be stabilized only by a signifivant amount of the negative GDD
($\approx$-9900 fs$^2$ in our case). However, the contribution of
higher-order dispersions can stabilize the pulse in the immediate
vicinity of zero GDD (black curves in Fig. \ref{fig5}). The pulse is
stable if even a part of spectrum is situated within the positive
GDD range (left picture in Fig. \ref{fig5}).

However, the NDR stability in the vicinity of zero GDD is very
sensitive to the spectral dependence of the losses. Presence of such
dependence leads to the multipulsing (harmonic mode-locking, gray
curves in Fig. \ref{fig5}). As it was found \cite{kalash3}, the
source of multipulsing is the insufficiently saturated net-gain,
which appears in the vicinity of zero GDD due to decrease of the
pulse energy caused by the spectral losses.

In analogy with the PDR, the spectral dependence of GDD can
initiate the chaotic mode-locking, when the negative GDD approaches
zero (gray curve in Fig. \ref{fig4}). The spectrum shifts from the
region, where the local minimum of GDD is located. This behaviour is reverse
in the comparison with that in the PDR. Thus,
the stability regions of PDR and NDR are disjointed by a region of
chaotic mode-locking in the vicinity of zero GDD.

\begin{figure}
   \begin{center}
   \vspace{-5 mm}
   \includegraphics[height=8cm]{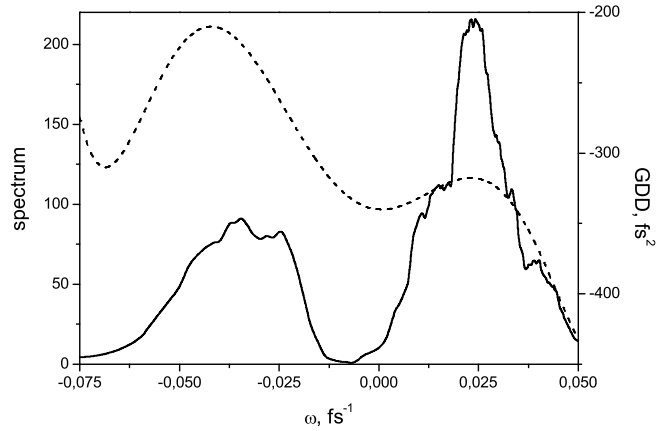}
   \vspace{-1 cm}
   \end{center}

   \caption[fig6]
   { \label{fig6}
GDD spectral profile (dashed) and corresponding simulated
  pulse spectrum in the negative dispersion regime with modulated dispersion and no spectral dependence of the losses.}
   \end{figure}

Just as the ``M-shaped'' spectra appear in the PDR, when the
fourth-order dispersion leads to the GDD growth on the spectrum
edges, the humps of the spectrum envelope exist in the NDR (Fig.
\ref{fig6}). In contrast to the PDR, the local spectral maxima
appear in the vicinity of local maxima of GDD.

\section{CONCLUSIONS}\label{s3}

We have studied stability conditions of a Cr:YAG oscillator in both, positive and negative dispersion regimes. In the NDR, the oscillator exhibits
strong tendency to harmonic mode-locking (i.e., multipulsing), which
can be suppressed only by a significant amount of the negative GDD. The GDD
value providing the multipulsing suppression depends on the GDD
shape, so that the soliton-like regime can exist even if the part of
the spectrum lies within the positive dispersion range. The spectrum
maximum shifts in the region, where the local maximum of GDD is
located. The presence of the spectrally-dependent losses enhances
the multipulsing in the vicinity of zero GDD. When the GDD is
frequency-dependent, the positive- and negative-dispersion
single-pulse regimes are disjointed by the GDD ranges with chaotic
and harmonic mode locking. In the PDR, approaching of the GDD to zero
results in a CW-amplification or a chaotic mode-locking. The
last regime appears if the GDD is frequency-dependent. The
stabilizing factor of such frequency dependence is the presence of
a local minimum (for the PDR) or maximum (for the NDR) of GDD in the
vicinity of the pulse spectrum.

\acknowledgments     
The work was supported by the Austrian National Science Fund (Fonds
zur F\"{o}rderung der Wissentschaftlichen Forschung (FWF), Projects
No. P20293 and P17973).

{}
\end{document}